\documentclass[10pt, onecolumn]{article}
\usepackage[left=1.0in, right=1.0in, top=1.0in, bottom=1.0in]{geometry}
\usepackage{graphicx,subfigure,epsfig}
\usepackage[verbose]{wrapfig}
\usepackage{float}
\usepackage{times}

\begin{document}
\title{Enhanced Light Emission from Erbium Doped Silicon Nitride in Plasmonic Metal-Insulator-Metal Structures}
\author{Yiyang Gong$^{1,*}$, Sel\c{c}uk Yerci$^{2}$, Rui Li$^{2}$, Luca Dal Negro$^{2}$, and Jelena Vu\v{c}kovi\'{c}$^{1}$ \\
	\small\textit{$^{1}$Department of Electrical Engineering, Stanford University, Stanford, CA 94305} \\
	\small\textit{$^{2}$Department of Electrical and Computer Engineering, Boston University, Boston, MA 02215} \\
\textit{*email: yiyangg@stanford.edu}
}
\maketitle
%\twocolumn[
%\begin{@twocolumnfalse}
\begin{abstract}
Plasmonic gratings and nano-particle arrays in a metal-insulator-metal structures are fabricated on an erbium doped silicon nitride layer. This material system enables simple fabrication of the structure, since the active nitride layer can be directly grown on metal. Enhancement of collected emission of up to 12 is observed on resonance, while broad off-resonant enhancement is also present. The output polarization behavior of the gratings and nano-particle arrays is investigated and matched to plasmonic resonances, and the behavior of coupled modes as a function of inter-particle distance is also discussed.
\end{abstract}
%\ocis{(240.6680) Surface Plasmons; (270.5580) Quantum electrodynamics}
%\end{@twocolumnfalse}
%]

\section{Introduction}
Erbium (Er)-doped materials have great potential as light sources in optoelectronics due to emission at the telecommunication wavelength of 1.54$\mu$m from the 4f level in erbium. In addition, Er doped (sub-stoichiometric) silicon oxide (SiO$_{x}$) and nitride (SiN$_{x}$) can be integrated with silicon complementary metal-oxide-semiconductor (CMOS) electronics, potentially serving as light sources for on-chip or off-chip communications. Because the 4f transition has a small oscillator strength, there have been many attempts to improve its emission efficiency for applications such as optical amplifiers. Recently, there has been significant work on the material properties of Er doped SiO$_{x}$ and SiN$_{x}$ to enhance the emission at 1.54$\mu$m. In particular, the material systems of Er sensitized by silicon nanoclusters in SiO$_{x}$ and SiN$_{x}$, Er doped SiN$_{x}$-Si superlattices, and Er doped amorphous silicon nitride (Er:SiN$_{x}$) have demonstrated an increase in emission and a reduction of non-radiative decay under optical and electrical pumping. \cite{Yerci_SNerb,Rli_SNerb,DalNegro_superlat,DalNegro_book,RLi_carriers,Wang_erb,Negro_el}.

In addition to improved material characteristics, emission of Er can also be increased by modifying the local electromagnetic density of states. In the weak coupling, or Purcell, regime, the spontaneous emission rate enhancement of an ensemble of emitters coupled to a cavity and emitting with free-space wavelength $\lambda$ (or frequency $\omega=2\pi c/\lambda$) is:
\begin{equation}
	F=\frac{3}{4\pi^2}\left(\frac{\lambda}{n}\right)^3\frac{Q}{V_{mode}}\overline{\psi(\theta,\vec{r},\omega)},
\label{eqn:Purcell}
\end{equation}
where $Q$ is the quality factor of the cavity mode, $V_{mode}$ is the mode volume of the cavity, and $\overline{\psi(\theta,\vec{r},\omega)}$ includes the average alignment of a dipole emitter with the cavity field and the decrease of enhancement for spatially and spectrally detuned emitters. Under assumptions of a randomly oriented dipole emitter and a Lorentzian cavity lineshape, $\psi(\theta,\vec{r},\omega)$ takes the form:
\begin{equation}
	\psi(\theta,\vec{r},\omega)=\frac{1}{3}\left(\frac{E(\vec{r})}{E_{max}}\right)^2\frac{(\omega/(2Q))^2}{(\omega-\nu)^2+(\omega/(2Q))^2)},
\label{eqn:Purcell1}
\end{equation}
where $E(\vec{r})$ is the spatial profile of the cavity mode and $\nu$ is the cavity frequency.

In particular, attempts to increase the spontaneous emission rate of erbium have employed photonic crystal cavities and plasmonic gratings \cite{Maria_erbcav,Polman_SPerb}. While photonic crystal cavities can achieve high quality ($Q$)-factors, their mode volumes have a lower bound of $(\lambda/(2n))^3$, which constrains the Purcell enhancement. Such enhancements are also limited to emitters near the spatial maximum of the cavity mode and spectrally coupled to the narrow cavity bandwidth, where the $(E(\vec{r})/E_{max})^2$ and Lorentzian spectral terms in Eqn. \ref{eqn:Purcell1} are respectively significant. Hence, large area and broadband enhancement are difficult to achieve. Furthermore, Purcell enhancement is limited by the quality factor of the emitter or the quality factor of the cavity, whichever is lower \cite{Woerdman_gamma}. The homogeneous linewidth of Er ions at room temperature is several nanometers, and thus limits the effective Purcell enhancement despite coupling to high-$Q$ cavities \cite{Desuvire_erbamp}. 

On the other hand, periodically patterned surface plasmon-polariton (SPP) modes have mode volumes that break the diffraction limit and have high $(E(\vec{r})/E_{max})^2$ in the active material. In addition, because the quality factor of the plasmonic modes is limited by ohmic losses of metal, plasmonic modes are broadband and can enhance emission over a large bandwidth. The coupling between SPP modes and a variety of materials, including quantum wells, colloidal quantum dots, silicon nanocrystals, and erbium, has been previously demonstrated, with spontaneous rate enhancement in several cases \cite{Polman_SPerb,Yablonovitch_SP,JV_Sim,Scherer_SP,Koichi_SP,Biteen_SiNC,YG_SiNC}.

The surface plasmon dispersion relation is given by \cite{Raether}:
\begin{equation}
    k=\frac{\omega}{c}\sqrt{\frac{\epsilon_{d}\epsilon_{m}(\omega)}{\epsilon_{d}+\epsilon_{m}(\omega)}},
\label{eqn:dispersion_eq}
\end{equation}
where $k=2\pi/\lambda$ is the wave-vector of the plasmon mode, $\omega$ is the mode frequency, $\epsilon_{d}=n^{2}$ is the dielectric constant of the dielectric material, $\epsilon_{m}(\omega)=1-(\omega_{p}/\omega)^2$ is the dielectric constant of the metal described by the Drude model, and $\omega_{p}$ is the plasma frequency of the metal. The decay length of the SPP mode in the dielectric material is:
\begin{equation}
    \kappa=\frac{\omega}{c}\sqrt{\frac{\epsilon_{d}^2}{|\epsilon_{d}+\epsilon_{m}(\omega)|}},
\label{eqn:decay_eq}
\end{equation}
which is in general comparable to $\lambda/n$ at small $k$-vectors. Because of this exponential decay into the active dielectric material, the enhancement of the spontaneous emission is limited. For example, silver nanoparticles deposited on an Er doped silicon-oxide sample enhanced the collected emission from an ensemble of Er ions by a factor of 2 \cite{Polman_SPerb}. In order to increase the emitter-field coupling, we propose to employ a metal-insulator-metal (MIM) device. Active MIM devices have been previously considered in only III-V semiconductors \cite{JV_Sim} and in organic thin films \cite{Gifford_MIM}. In addition, MIM waveguides with sputtered oxide as a passive insulator have been previously demonstrated, with extreme sub-wavelength dimensions \cite{Miyazaki_MIM}, and coupling of emitters to MIM modes have also been theoretically considered \cite{Brongersma_MIMtheory}. The dispersion relation of the MIM mode with semi-infinite metal thicknesses, insulator thickness $t$, and field symmetric along the direction perpendicular to the insulator layer, is given by the transcendental equation \cite{Kurokawa_MIM}:
\begin{equation}
    -\frac{k_{m}}{k_{d}}=\frac{\epsilon_{m}(\omega)}{\epsilon_{d}} \mbox{tanh}(k_{d}t/2),
\label{eqn:MIM_eq}
\end{equation}
where $k_{m}=\sqrt{k^2-\epsilon_{m}(\omega)(\omega/c)^2}$ and $k_{d}=\sqrt{k^2-\epsilon_{d}(\omega/c)^2}$ are defined in terms of the propagation k-vector $k$, and $t$ is the thickness of the insulator layer. Because the two metal surfaces confine the SPP mode from both sides of the active material, the mode volume of the MIM mode is decreased relative to that of the SPP mode supported by metal on only one side of the insulator. In addition, the symmetric field mode of a MIM structure has higher overlap with the active material than the mode of a single sided design. The material properties of the Er doped nitride layer are also advantageous, as amorphous silicon nitride is a very robust material and can be deposited on a gold substrate at room temperature. Thus, the MIM structure with Er:SiN$_{x}$ as the insulator can be fabricated with a bottom-up procedure, without the need for complicated epitaxial lift-off processes as in the case of crystalline semiconductors \cite{JV_Sim}.

\begin{figure}[hbtp]
\centering
\includegraphics[width=5.0in]{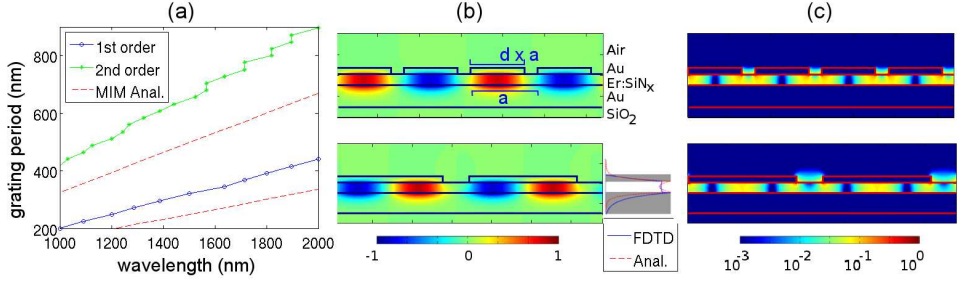}
\caption{(a) The wavelengths of the first and second order field-symmetric MIM SPP modes for different grating periods, with a 52nm thick layer of Er:SiN$_{x}$ between the metal layers. The analytical solution for a MIM structure with semi-infinite metal thickness and the same Er:SiN$_{x}$ thickness is also shown. (b) The magnetic field ($B$) and (c) the electric field intensity ($|E|^2$) of the first (top) and second (bottom) order modes near free space $\lambda_{0}\approx 1500$nm. The MIM SPP modes in a 2D simulation have $B$ fields perpendicular to plane of the figures, while the $E$ field is restricted to the plane of the figures. The inset of (b) shows the magnetic field through a vertical slice of the structure, as well as the analytical solution of a MIM system with semi-infinite metal thicknesses and a 52nm Er:SiN$_{x}$ spacer layer \cite{Kurokawa_MIM}.}
\label{fig:modes}
\end{figure}

First, we analyze the grating pattern using the two dimensional Finite Difference Time Domain (2D-FDTD) method with parameters in our previous work \cite{YG_SiNC,YG_sim}. The structure has a 100nm gold substrate, with plasma frequency $\omega_{p}=2\pi c/(160\times 10^{-9}\mbox{m})$, a 52nm Er:SiN$_{x}$ layer with index of refraction $n$=2.2, and a 30nm thick top gold grating (Fig. \ref{fig:modes}(b)). In particular, we fix the grating duty cycle (the ratio between the gold bar width and the grating period, as in Fig. \ref{fig:modes}(b)) to be $d=0.8$, and vary the grating period. We find the wavelengths of the band edge SPP modes as a function of grating period, as plotted in Fig. \ref{fig:modes}(a). We focus on the SPP MIM mode that has symmetric field (and anti-symmetric charge) along the growth direction (up-down in Fig. \ref{fig:modes}(b)-(c)), as it will have high overlap with the active material. In addition, we focus only on the band edge modes in the plasmonic band diagram, which support integer multiples of the SPP half-wavelengths within the grating period and exhibit the strongest emission enhancement \cite{YG_SiNC}. The mode that fits $p$ SPP half-wavelengths in the grating period is labeled as the p$^{th}$ mode. Because the Er emission wavelength is much longer than the plasma wavelength of gold, the structures operate in the ``linear" regime of the SPP dispersion relation, where the band edge modes' wavelengths are approximately linear with respect to the grating period. We examine the magnetic field ($B$) and electrical field intensity ($|E|^2$) of the first and second order modes, where the first order mode (Fig. \ref{fig:modes}(b)-(c), top) is at the $X$-point of the dispersion relation (wavevector $k=\pi/a$, where $a$ is the grating period), while the second order mode (Fig. \ref{fig:modes}(b)-(c), bottom) is at the $\Gamma$ point ($k=0$). The solution of Eqn. \ref{eqn:MIM_eq} for the MIM mode with semi-infinite metal thicknesses and the same Er:SiN$_{x}$ parameters is also plotted in Fig. \ref{fig:modes}(a) for the same periodicities, and the analytical magnetic field profile of such a MIM mode is shown in the inset of Fig. \ref{fig:modes}(b). The dispersion relation of the grating is blue shifted from the analytical model of Eqn. \ref{eqn:MIM_eq}, as the MIM modes for the grating are no longer fully confined and slightly overlaps with air.

\section{Enhancement of photoluminescence via MIM modes}
\begin{figure}[hbtp]
\centering
\includegraphics[width=4.0in]{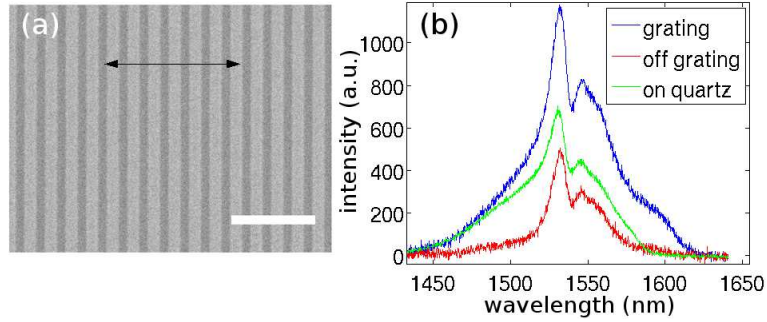}
\caption{(a) Top gold grating imaged by SEM. The marker denotes 2 $\mu$m, while the double arrows indicate the approximate alignment of the polarizer ($0^{\circ}$) (b) PL from on and off the grating structure, as well as from a reference sample with the same Er:SiN$_{x}$ thickness grown on quartz. The same excitation power was used in all three cases. The region off the grating is a region with Er:SiN$_{x}$ on top of Au without the top metal layer.}
\label{fig:structure}
\end{figure}

The structures under study are fabricated by first evaporating an 8nm Chromium (Cr) sticking layer followed by a 100nm gold layer on top of an oxidized silicon wafer. The 1$\mu$m oxidized silicon serves as a diffusion blocking layer beneath the gold. Then, a 52nm erbium doped amorphous silicon nitride (Er:SiN$_{x}$) layer is deposited on top of the gold substrate by N$_{2}$ reactive magnetron co-sputtering from Si and Er targets in a Denton Discovery 18 confocal-target sputtering system \cite{Yerci_SNerb,Rli_SNerb}. The growth is followed by a post annealing process in a rapid thermal annealing furnace at 600$^{\circ}$C for 600s under forming gas (5\% H$_{2}$, 95\% N$_{2}$) atmosphere. While the optimal annealing temperature to maximize erbium emission efficiency is 1150$^{\circ}$C, degradation of the gold substrate at high temperatures restricts the annealing temperature to be below 600$^{\circ}$C. The top side grating pattern is then formed by electron beam lithography with polymethylmethacrylate (PMMA) as the resist. Finally, the grating is fabricated by depositing a 3 nm Cr sticking layer followed by a 30nm layer of Au on top of the patterned resist, and lifting off in acetone. The final structure is shown in Fig. \ref{fig:structure}(a), imaged in a scanning electron microscope (SEM).

%fig1.m from 20090713

In the experiment, we pump the MIM structure through the top-side plasmonic grating with a 400nm laser diode, focused by a 100X objective lens with numerical aperture NA=0.5. The beam is focused down to approximately a 3$\mu$m radius, smaller than the size of the structures (10$\mu$m $\times$ 10$\mu$m). The output is collected through the same objective, and is directed to a spectrometer with the pump laser filtered out. The output photoluminescence (PL) for Er:SiN$_{x}$ deposited on gold, as well as for a sample grown and annealed on a quartz substrate under the same conditions,  is shown in Fig. \ref{fig:structure}(b). The PL spectrum from the Er:SiN$_{x}$ layer on gold is similar to the spectrum of Er:SiN$_{x}$ deposited on quartz, while the total integrated intensity from the sample on the gold substrate is reduced by approximately a factor of 2. We perform lifetime measurements using the demodulation technique \cite{demod}, and find that the Er:SiN$_{x}$ deposited on gold has a lifetime of 100 $\mu$s. We also measure the lifetime of the quartz substrate sample, both with and without a 30nm gold layer on top of the Er:SiN$_{x}$ layer, and observe lifetimes of 100-130 $\mu$s. On the other hand, for samples of Er:SiN$_{x}$ of the same thickness grown on quartz annealed at 1100$^{\circ}$C,  we observe lifetimes of 1.0ms and 1.5ms with and without a thick gold layer on top, respectively. In addition, the integrated PL for such samples were approximately 3 times higher than any of the samples annealed at lower temperatures. Based on these measurements, we conclude that the large decrease in Er emission efficiency is due to non-optimal annealing conditions. However, we also attribute some losses to the imperfect film quality on top of the gold substrate, but not to non-radiative decay mediated by SPP modes on the gold substrate.

\begin{figure}[hbtp]
\centering
\includegraphics[width=4.5in]{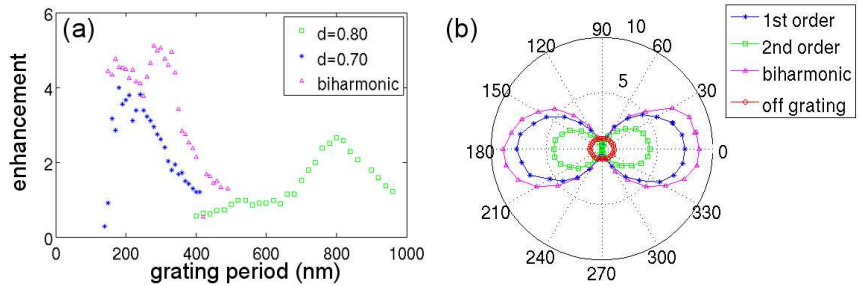}
\caption{(a) The enhancement of integrated emission from grating structures relative to off grating areas, as a function of grating period. The three curves correspond to two different duty cycles ($d$), and to a biharmonic grating. (b) The enhancement of emission as a function of polarization angle for the resonant first and second order grating modes, as well as for the resonant biharmonic grating. The angle dependence of the Er:SiN$_{x}$ on Au off grating is also shown.}
\label{fig:oneDPL}
\end{figure}
%data is 2009/06/23 run10 - biharm
%data is 2009/07/21 run1 - biharm angle
%data is 2009/07/13 runs have single and double, including angled data
% anglefigs.m in 20090716

Next we vary the period of the grating, and observe the enhancement of PL relative to the sample without the top metal layer (labeled as ``off grating"). As seen in the PL trace from an example grating structure in Fig. \ref{fig:structure}(b), the output is enhanced relative to the case without the grating. Because the observed plasmonic mode linewidth is broader than the erbium emission, we plot the integrated PL intensity over the emission spectrum for different periods of the grating, normalized by PL from an area off grating (Fig. \ref{fig:oneDPL}(a)). The duty cycle $d$=0.7, was fabricated at low grating periodicities, due to minimum feature sizes required for liftoff, while a higher duty cycle of $d$=0.8 was fabricated for higher grating periods. We find several resonant grating periods with maximum enhancement of approximately 4 and 3, respectively, for the periods of 250nm and 800nm. These correspond well to the FDTD calculated grating periods of 300nm and 740 for the first and second order modes with resonances around $\lambda_{0}=1.54\mu$m. The particular resonant grating period of 250nm is well below the $\lambda/(2n)$ cutoff of purely dielectric modes, suggesting that the enhanced emission is coupled to plasmonic modes. Finally, the enhancement rolls off as the grating becomes non-resonant, with the enhancement of emission returning to unity. From the FDTD calculation, we estimate the $Q$-factors of the first and second order modes to be 3 and 5, respectively. We do not expect the linewidth of these modes to be broadened by collecting more than one point of the dispersion relation, as we filter for one emission direction in the Fourier plane of the collection path without observing any changes in the spectral shape of the emission.

In addition, we perform polarization measurements on the periodic array by placing a half waveplate followed by a polarizer in the collection path. In all subsequent measurements, the horizontal direction (0$^{\circ}$) corresponds to the polarizer set perpendicular to the grating bars (i.e., aligned with the wave-vector $k$), as shown in Fig. 1(a). In Fig. \ref{fig:oneDPL}(b), we plot the angle dependence of the emission from structures at the two maxima of enhancement in Fig. \ref{fig:oneDPL}(a), as well as from a region off grating. First, we find that the emission from off grating  is unpolarized, as the angle dependence of the emission is mostly flat with minor perturbation coming from the angle dependent deflection from the optics. For the first order mode with the highest intensity enhancement, we find that the 0$^{\circ}$ polarization is greatly enhanced compared to the off grating emission, by factor of 7. As the polarization dependent enhancement rejects one half of the unpolarized off grating emission, the peak angular enhancement corresponds well to the peak enhancement of 3 to 4 measured without the polarizer. The maximum enhancement at 0$^{\circ}$ is expected, as the SPP mode is a longitudinal mode with polarization parallel to its $k$-vector. On the other hand, the emission in the orthogonal polarization is greatly suppressed. The suppression indicates that the PL in the orthogonal direction does not couple to the grating mode. In such a manner, the one-dimensional grating is an efficient polarization selective enhancement device.

\begin{figure}[hbtp]
\centering
\includegraphics[width=4.5in]{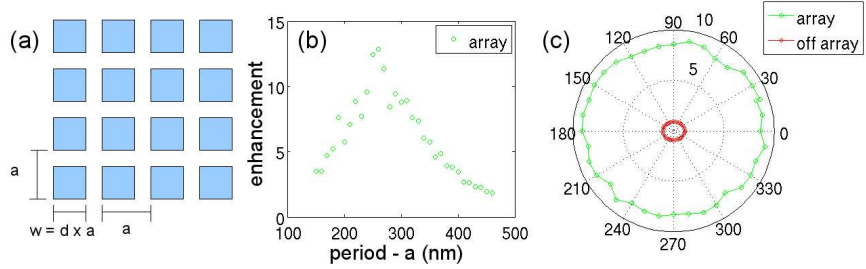}
\caption{(a) The design of the square array of square metal particles. (b) The enhancement of emission from the shown nano-particle array structures relative to off-array areas, as a function of array period ($a$). (c) The angle dependence of the output from a near-resonant nano-particle array with lattice constant of $a$=290nm.}
\label{fig:arrayPL}
\end{figure}

% 2009/07/13 - run6 - array (fig3.m)
% 2009/07/16 - array angle

We also attempt to enhance the output by employing biharmonic gratings as in previous reports \cite{YG_SiNC}. Again, due to limitations in fabrication, the duty cycle was kept low for proper liftoff. The maximum duty cycle fabricated has average metal coverage of 56\%. The enhancement with respect to the fundamental grating period is plotted in Fig. \ref{fig:oneDPL}(a) and the angle dependence of the emission from the grating with maximum enhancement plotted in Fig. \ref{fig:oneDPL}(b). As expected, the resonant grating period matches well with the resonance of single period grating, and also maintains the broad resonance of the first order grating modes. However, the maximum angle dependent enhancement increased to only 9. 

%This is expected, as the grating periods supporting the first order modes should be below the light-line and thus should not have reached the collection optics. However, the observed enhancement of emission from such modes suggests that the grating was sufficiently rough to extract the emission via scattering, or that single particle resonances are observed. In either case, further manipulation of the dispersion relationship would not greatly increase the extracted emission.

In order to enhance the output for both polarizations, we fabricate nano-particle arrays using the same procedure. In particular, we design a square lattice of square metal particles, with similar lattice constant as the one-dimensional gratings and duty cycle $d$=0.7 as defined in Fig. \ref{fig:arrayPL}(a). We pump these structures in the same manner and observe the intensity of the output, plotted in Fig. \ref{fig:arrayPL}(a). We note a maximum enhancement of approximately 12 for some structures, and the resonant lattice constant is similar to the resonant grating period of the one-dimensional grating. In addition, by measuring the angular dependence of the enhancement for the resonant lattice constant $a$=290nm as shown in Fig. \ref{fig:arrayPL}(b), we notice that the output is unpolarized, which is expected from the symmetry of the two orthogonal plasmonic modes in the square lattice. We do not observe the diagonal traveling modes with grating period 250nm/$\sqrt{2}$, most likely due to the lack of symmetry of the individual square particles.

%Finally, we attempt time-resolved measurements using the demodulation method shown in previous works. In particular, the pump beam is chopped from 0-2000 Hz, and the output signal is integrated by a lock-in amplifier for 3s at the first harmonic. Emitters with faster lifetimes decay at a faster rate, and thus reach the zero emission level faster, than emitters with slower lifetimes, for the same chopping frequency. As the chopping frequency is increased, emitters with faster lifetimes will maintain a higher amplitude at the first harmonic that emitters with slower lifetimes. Assuming a infinitely fast rise time, the demodulation as a function of the chopping frequency is:

%We take several traces over sample with gratings and without gratings, and plot the normalized traces in Figure. The bulk sample emission, with a lifetime of 100us, clearly shows more demodulation than emission from the gratings, which have fitted lifetimes of 50us. In addition, we find the lifetimes of several gratings, and plot them against the PL enhancement factors. We first note that at the grating period with the highest enhancement, there is a local minimum in the emission lifetime. We see that the lifetime recovers to the bulk values for non-optimal grating periods. 

\section{Role of grating and particle resonances}
\begin{figure}[hbtp]
\centering
\includegraphics[width=4.5in]{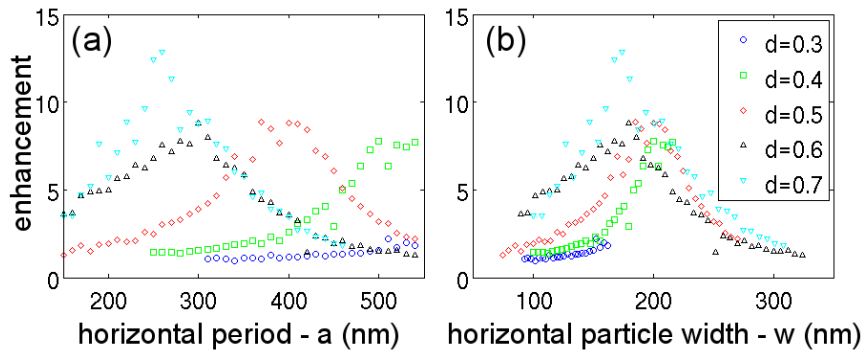}
\caption{(a) The enhancement of emission from a square array of square metallic particles (Fig. \ref{fig:arrayPL}(a)), as the duty cycle ($d$) and lattice constant ($a$) are varied. (b) The same set of data is replotted against the particle width, $w=d\times a$.}
\label{fig:arrayduty}
\end{figure}

As shown in Figs. \ref{fig:oneDPL} and \ref{fig:arrayPL}, we managed to collected emission from the first order modes of both the grating and the nano-particle array although they should be positioned below the light line. This suggests that scattering from rough metal surfaces or plasmonic resonances localized to individual metal particles, as opposed to coupled grating modes, may be the dominant effect in the observed enhancement. In order to study this effect in greater detail, we fabricate square arrays with the design in Fig. \ref{fig:arrayPL}(a), but vary the duty cycle from $d$=0.3 to $d$=0.7. We plot the enhancements corresponding to various duty cycles in Fig. \ref{fig:arrayduty}(a). By plotting the enhancement versus the particle width ($w = d\times a$) in Fig. \ref{fig:arrayduty}(b), we notice that the resonances correspond to similar particle widths, especially for low duty cycle. This indicates that modes localized to individual particles are the main contributors to emission enhancement and extraction. Furthermore, as the duty cycle increases, the particle width corresponding to maximum enhancement decreases. This is explained by the formation of coupled modes between the particles. For coupled modes, as shown by the field profiles in Fig. \ref{fig:modes}(b)-(c), the MIM mode is more confined to the Er:SiN$_{x}$ layer, thus increasing the effective index. Such an effect in turn slightly reduces the frequency of the MIM mode for the same particle width, and requires a smaller particle to achieve the same resonant wavelength. We also note that the arrays of smaller period maintain relatively high enhancement despite being away from the plasmonic resonance. However, the modes of these closely packed particles resemble traveling waves of an unpatterned MIM slab (instead of individual particle resonances) and thus have wide bandwidth. Finally, the maximum enhancement increases with duty cycle. Because the effective pump reaching the Er decreases with increasing duty cycle (due to the presence of more metal), this indicates that the actual enhancements in the high duty cycle arrays are even higher than 12 and higher than in the low duty cycle case. 

%In addition, we observe that the enhancement peaks for certain grating periods, which varies greatly as the duty cycle of the nano-particle array is changed. However, if we plot the particle size by multiplying the particle array periodicity by the duty cycle, we observe that the highest enhancements all come around the same particle size. Such evidence suggests that the enhancement comes from an MIM mode that is confined to individual particles. 

\begin{figure}[hbtp]
\centering
\includegraphics[width=5.0in]{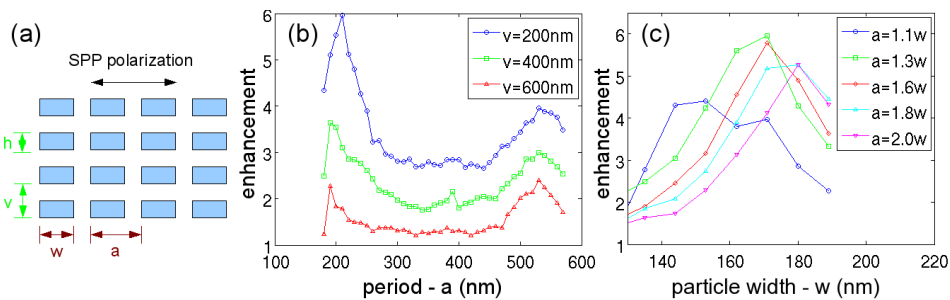}
\caption{(a) Asymmetric particle arrays used to examine the role of inter-particle distance in the directions parallel and perpendicular to the plasmonic mode polarization. The vertical period $v$ and horizontal period $a$ are changed independently of each other. The polarization of the plasmonic resonances observed is shown by the double arrows. (b) The enhancement of emission from asymmetric particle arrays as a function of horizontal period $a$, while $v$ is changed. Vertical period v is increased in increments of 200nm. The particle has a width of w=0.8a, and height of h=0.4v for all measurements. The horizontal period $a$ that produces maximum enhancement remains the same for all $v$. (c) The enhancement of emission from arrays where $v$ is fixed at 600nm, $h$ is fixed at 360nm, and the horizontal width and period are varied independently.}
\label{fig:varyspace}
\end{figure}

Finally, we attempt to quantify the coupling between the particles as a function of particle separation and polarization. To do this, we fabricate rectangular arrays, where in the horizontal direction, gratings have width $w=0.8a$, and we vary the period $a$. In the other direction, we vary period ($v$) by increments of 200nm, with metal bar height ($h$) fixed to be 0.4$v$ (Fig. \ref{fig:varyspace}(a)). We measure the enhancements for different horizontal and vertical distances without a polarizer, and plot the results in Fig. \ref{fig:varyspace}(b). Here, we observe that the resonant horizontal periods of 210nm and 550nm are the same for all three vertical periods, while the overall enhancement is reduced for higher vertical period. This suggests that the plasmonic oscillations that enhance emission at the resonant grating periods are primarily in the horizontal direction, and that the oscillation frequency does not change with inter-particle separation in the vertical direction. The reduction in enhancement for larger vertical separation is due to reduced overlap with the active material and the off-resonant particle height for large $v$.

%varyspace.m in 20090713

In order to determine the role of inter-particle spacing in the direction of the plasmonic mode polarization, we fabricate gratings with the same fixed $v$=600nm and $h$=360nm, but vary the horizontal particle width $w$ and horizontal period $a$ (Fig. \ref{fig:varyspace}(a)). We plot the enhancement against the particle width for different horizontal separations in Fig. \ref{fig:varyspace}(c). As with the square lattice arrays, we observe a distinct shift in the resonant horizontal particle width as the horizontal inter-particle spacing is varied. The resonant particle width increases as the particles are pulled away from each other, again suggesting a decrease in the coupling between SPP modes on separate particles and a decrease in the overlap with the active material. This confirms that the particle arrays have resonant fields confined to the individual particles for large inter-particle separation, while supporting coupled modes for small inter-particle separation. We also note that the peak enhancement decreases with increasing particle separation above $a \ge 1.3w$, suggesting that coupled modes increase the overall enhancement. Such evidence corroborates the increased overlap between coupled modes and the active material. While the $a=1.1w$ case has lower peak enhancement, pumping efficiency and outcoupling efficiency are reduced because of large metal coverage, and effective enhancement may be larger.

\section{Conclusions}
We have demonstrated an order of magnitude increase in collected emission intensity from a MIM structure with Er:SiN$_{x}$ as the active material. The device has been fabricated with simple bottom-up procedure and does not require complicated flip-chip bonding procedures. In addition, we have identified both local and coupled SPP modes supported by metallic gratings and nano-particle arrays.

The device could be improved significantly by improving the material properties of the Er:SiN$_{x}$. First, the internal efficiency can be improved by reducing the large non-radiative losses in the Er:SiN$_{x}$ layer. This could be done by improving the growth process on top of a metal substrate, or designing devices that encapsulate higher temperature annealed Er:SiN$_{x}$ with metal. In addition, the MIM structure certainly has the potential to be electrically pumped since the metal layers could serve as the electrodes, and electrical pumping of silicon rich nitride layers has been investigated \cite{Negro_el}.

The authors would like to acknowledge the MARCO Interconnect Focus Center, the AFOSR under MURI Award No. FA9550-06-1-0470, the Toshiba Corporation, and the NSF GRF for funding. The fabrication has been performed in the Stanford Nanofabrication Facilities.

\end{document}